\documentclass[prb,twoside,twocolumn,10pt,floatfix,showpacs,citeautoscript,superscriptaddress]{revtex4-2}

\usepackage{amsmath}
\usepackage{amssymb}
\usepackage{booktabs}
\usepackage{comment}
\usepackage{glossaries}
\usepackage{graphicx}
\usepackage[version=4]{mhchem}
\usepackage[per-mode=symbol]{siunitx}
\usepackage{tabularx}
\usepackage[dvipsnames]{xcolor}

\usepackage[colorlinks=true]{hyperref}
\newcommand \Title{
    Understanding correlations in \texorpdfstring{\ce{BaZrO3}}{BaZrO3}:Structure and dynamics on the nano-scale
}

\hypersetup{
    pdfauthor={E. Fransson, P. Rosander, P. Erhart, G. Wahnström},
    pdftitle={\Title},
    pdffitwindow=false,
    pdfstartview={FitH},
    pdfnewwindow=true,
    colorlinks=true,
    linkcolor=black,
    citecolor=black,
    filecolor=black,
    urlcolor=black,
}

\graphicspath{{figures/}}

\setacronymstyle{long-short}
\newacronym{dft}{DFT}{density functional theory}
\newacronym{fwhm}{FWHM}{full width half maximum}
\newacronym{md}{MD}{molecular dynamics}
\newacronym{mlp}{MLP}{machine learning potential}
\newacronym{nep}{NEP}{neuroevolution potential}
\newacronym{scp}{SCP}{self-consistent phonon}
\newacronym{cx}{CX}{van-der-Waals-density functional with consistent exchange}
\newacronym{xrd}{XRD}{X-ray diffraction}

\newcommand{\gpumd}{\textsc{gpumd}}
\newcommand{\dynasor}{\textsc{dynasor}}

\newcommand{\phonopy}{\textsc{phonopy}}
\newcommand{\ovito}{\textsc{ovito}}

\DeclareSIUnit\fu{\text{f.u.}}

\DeclareSIUnit\angstrom{\text {Å}}
\renewcommand{\vec}[1]{\ensuremath\boldsymbol{#1}}
\renewcommand{\epsilon}[0]{\varepsilon}

\newcommand{\addchalmers}{
    Department of Physics,
    Chalmers University of Technology,
    SE-41296, Gothenburg, Sweden
}

\begin{document}

\title{
Understanding correlations in \texorpdfstring{\ce{BaZrO3}}{BaZrO3}: Structure and dynamics on the nano-scale
}

\author{Erik Fransson}
\author{Petter Rosander}
\author{Paul Erhart}
\author{G\"oran Wahnstr\"om}
\email{goran.wahnstrom@chalmers.se}
\affiliation{\addchalmers}

\begin{abstract}
Barium zirconate \ce{BaZrO3} is one of few perovskites that is claimed to retain an average cubic structure down to \qty{0}{\kelvin} at ambient pressure, while being energetically very close to a tetragonal phase obtained by condensation of a soft phonon mode at the R-point.
Previous studies suggest, however, that the local structure of \ce{BaZrO3} may change at low temperature forming nanodomains or a glass-like phase.
Here, we investigate the global and local structure of \ce{BaZrO3} as a function of temperature and pressure via molecular dynamics simulations using a machine-learned potential with near density functional theory (DFT) accuracy.
We show that the softening of the octahedral tilt mode at the R-point gives rise to weak diffuse superlattice reflections at low temperatures and ambient pressure, which are also observed experimentally.
However, we do not observe any \emph{static} nanodomains but rather soft \emph{dynamic} fluctuations of the \ce{ZrO6} octahedra with a correlation length of 2 to \qty{3}{\nano\meter} over time-scales of about \qty{1}{\pico\second}.
This soft dynamic behaviour is the precursor of a phase transition and explains the emergence of weak superlattice peaks in measurements.
On the other hand, when increasing the pressure at \qty{300}{\kelvin} we find a phase transition from the cubic to the tetragonal phase at around \qty{16}{\GPa}, also in agreement with experimental studies.
\end{abstract}

\maketitle

\section{Introduction}

Perovskite oxides constitute a prominent class of materials with a wide range of different properties, such as ferroelectricity, colossal magnetoresistance, electronic and/or ionic conductivity,  piezoelectricity, superconductivity, metal-insulator transition, luminescence, and many more \cite{Bhalla2000}.

The prototypical oxide perovskite structure is cubic, with the general chemical formula \ce{ABO3}, where the A and B sites can accommodate a wide variety of elements from the periodic table. 
Many perovskites are cubic at high temperatures but upon cooling most undergo one or several structural phase transitions, which depend sensitively on the choice of A and B.
These phase transitions are often related to tilting of the \ce{BO6} octahedra, typically referred to as antiferrodistortive transitions.
Commonly they are out-of-phase and in-phase tilting phonon modes related to instabilities at the R and/or M-points of the Brillouin zone.

Barium zirconate \ce{BaZrO3} is rather unique among the oxide perovskites. 
Neutron powder diffraction studies show that \ce{BaZrO3} at ambient pressure maintains its high temperature cubic structure down to temperatures close to zero Kelvin \cite{Akbarzadeh2005, Knight2020, PerJedRomPioLinHylWah20}.
While the antiferrodistortive R-tilt mode softens substantially with decreasing temperature, its frequency remains positive as the temperature approaches zero Kelvin \cite{Zheng2022,RosFraBrauTouBouAndBosMaeWah23}.

While the latter experiments have clearly established the long-range order, the short-range order of the cubic \ce{BaZrO3} phase is more controversial. 
Raman spectra show pronounced peaks despite that first-order scattering is prohibited by symmetry reasons for cubic systems \cite{Chemarin2000, Karlsson2008, Giannici2011}.
This has been interpreted as evidence for distorted nanodomains with lower than cubic symmetry, giving rise to first-order broad Raman spectra \cite{Chemarin2000, Lucazeau2003}.
A somewhat similar idea, an ``inherent dynamical disorder'', has been put forward to account for the apparent local deviation from the cubic structure identified by Raman spectroscopy \cite{Giannici2011}.
On the other hand, Raman studies of \ce{BaZrO3} single crystals associated these spectral features to second-order events, but stated that it is likely that the overall scattering intensity finds its origin in some other type of local disorder \cite{Toulouse2019}.
It has also been argued that a structural ``glass state'' may be formed upon cooling due to the extremely small energy differences between the phases allowed from condensation of the R mode \cite{Lebedev2013}.
The structural order could then be distorted on the local scale but appear cubic in diffraction experiments.

Recent electron diffraction experiments by Levin \textit{et al.} \cite{Levin2021} suggest that \ce{BaZrO3} undergoes a local structural change associated with correlated out-of-phase tilting of the \ce{ZrO6} octahedra when the temperature is reduced below \qty{80}{\kelvin}.
They found weak, but clear, diffuse scattering intensity at the R-point $(3/2, 1/2, 1/2)$, where the soft mode connecting the cubic to the tetragonal phase is located; yet their average structure remained cubic.
The authors suggested that nanometer-sized domains (``nanodomains'') with local tetragonal structure could explain the diffraction results.
The size of these domains was estimated to be about 2 to \qty{3}{nm} based on the \gls{fwhm} of the diffraction peaks.
They stated that the emergence of these relatively sharp superlattice reflections resembles a phase transition more than dynamic correlations, but their measurements could not conclusively discern between static and dynamic effects.

The pressure dependence of \ce{BaZrO3} at room temperature has been investigated by several authors \cite{Yang2014, Chemarin2000, Gim2022, Toulouse2022}.
In a recent combined X-ray diffraction and Raman spectroscopy study \cite{Toulouse2022}, it was found that \ce{BaZrO3} undergoes a single phase transition around \qty{10}{\GPa} from the cubic ($Pm\overline{3}m$) to the tetragonal ($I4/mcm$) phase and retains that structure up to \qty{45.1}{\GPa}. 
No second phase transition to an orthorhombic or any other tilted phase was observed.
This confirms a previous high pressure X-ray diffraction study from 0 to \qty{46.4}{\GPa} \cite{Yang2014}, where also a single transition from the cubic to the tetragonal phase was obtained, but at the considerably higher pressure of \qty{17.2}{\GPa}.
However, a recent study based on Raman spectroscopy \cite{Gim2022} found two structural phase transitions, the first from the cubic to a rhombohedral ($R\overline{3}c$) phase at \qty{8.4}{\GPa} and the second from the rhombohedral to the tetragonal phase at \qty{11}{\GPa}.

Here, we construct a machine-learned potential using the \gls{nep} approach trained with \gls{dft} data to be able to simulate the system over long time-scales (\qty{100}{ns}) using large systems (15 million atoms) with near \gls{dft} accuracy.
The phase diagram for \ce{BaZrO3} is mapped out as a function of temperature and pressure and compared with experiments. 
The static and dynamic structure factors as a function of wavevector and frequency are computed and their dependence on temperature and pressure are investigated.
Detailed and direct comparison is made with the electron diffraction data by Levin \textit{et al.} \cite{Levin2021} and the dynamics close to the R-point is clarified.
Finally, the spatial and temporal correlations of the local tilt angles for each individual \ce{ZrO6} octahedron are computed to elucidate the three-dimensional structure and dynamics of \ce{BaZrO3} as a function of temperature and pressure.

\section{Results}

\subsection{Instabilities and phase diagram}

\begin{figure}[ht]
\centering
\includegraphics{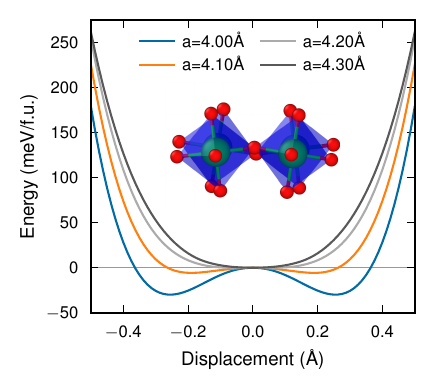}
\caption{
    The potential energy landscape for the R-tilt mode obtained with the \gls{nep} model as function of the oxygen atom displacement.
    The inset shows the atoms in a tilted structure, where red is oxygen, green zirconium and the blue faces show the \ce{ZrO6} octahedra.
}
\label{fig:Rmode_landscape}
\end{figure}

\begin{figure*}[ht]
\centering
\includegraphics{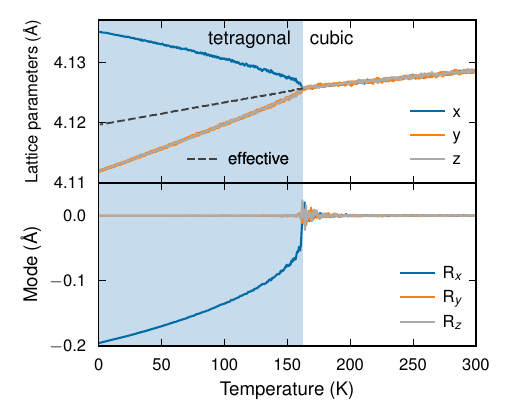}
\includegraphics{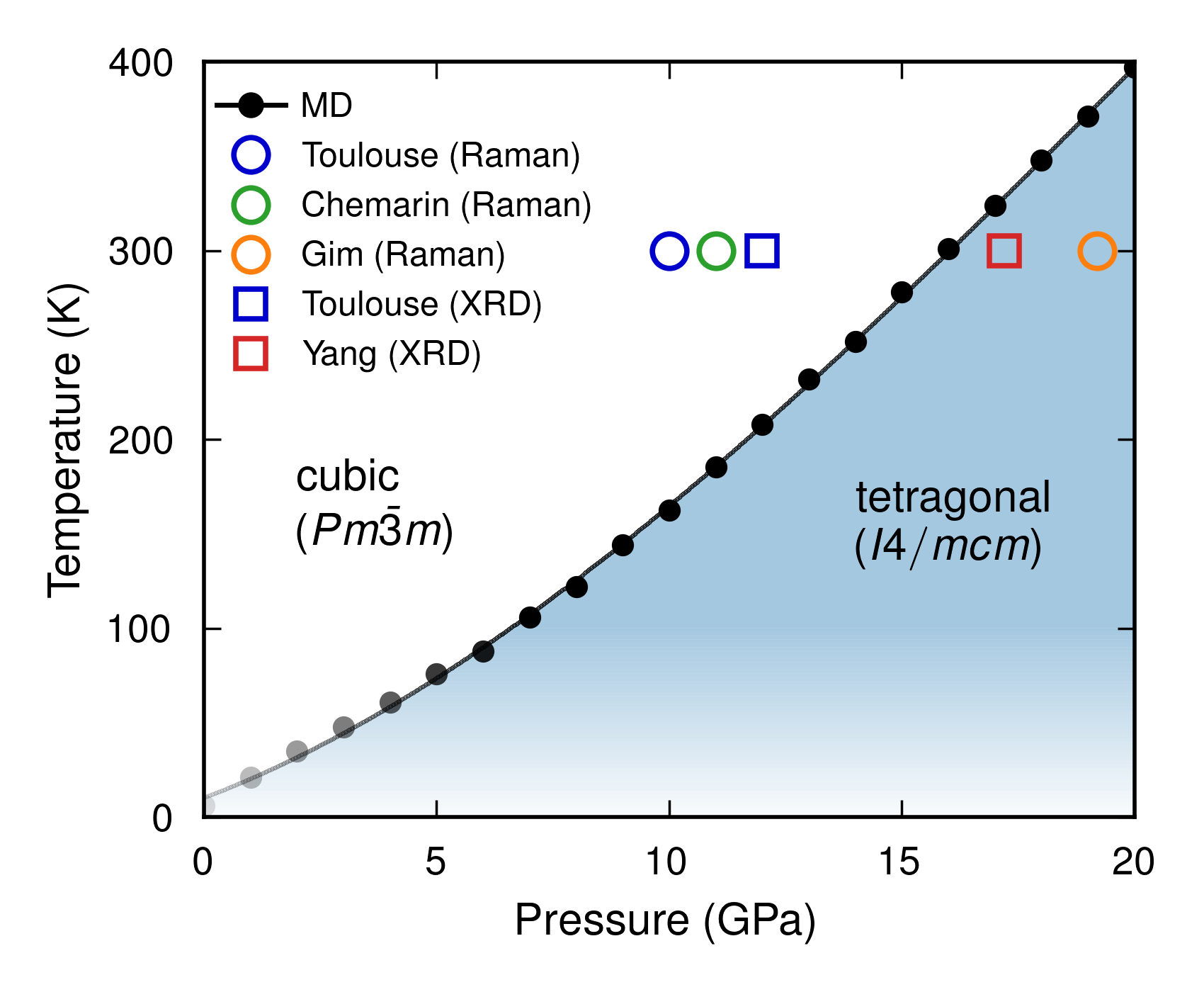}
\caption{
    a) Temperature dependence of lattice parameters and R-mode coordinates at \qty{10}{\giga\pascal} from a cooling \gls{md} run in the NPT ensemble.
    The phase transition from the cubic to the tetragonal phase is observed at about \qty{160}{\kelvin}. The effective lattice parameter for the tetragonal phase is given by $V^{1/3}$, where $V$ is the volume per formula unit.
    b) Phase diagram from cooling \gls{md} runs. 
    Here, the region below \qty{100}{\kelvin} is shown with increasing transparency to reflect the uncertainty due to the classical sampling in \gls{md}.
    Experimental Raman spectroscopy data from Refs.~\citenum{Chemarin2000,Toulouse2022,Gim2022} and \gls{xrd} data from Refs.~\citenum{Yang2014,Toulouse2022}.
}
\label{fig:phase_diagram}
\end{figure*}

\Gls{dft} calculations based on the CX functional yield a lattice parameter for cubic \ce{BaZrO3} of \qty{4.20}{\angstrom}, for which the phonon dispersion curves show only a very weak instability at the R-point \cite{PerJedRomPioLinHylWah20}. 
When decreasing the lattice parameter the instability at the R-point increases and for \qty{4.00} {\angstrom} the dispersion curves also show an instability at the M-point (\autoref{sfig:dispersion}). 

In \autoref{fig:Rmode_landscape} we show the static energy landscape along the R-tilt mode as function of the oxygen atom displacement.
For \qty{4.00}{\angstrom} a clear double well energy landscape is obtained with depths equal to \qty{-29.8}{\milli\electronvolt\per\fu} and located at \qty{+-0.25}{\angstrom}.
This corresponds to a tilt angle of \qty{7.1}{\degree}.
We note that the M-mode instability for \qty{4.00}{\angstrom} is barely visible on the same energy scale (\autoref{sfig:mode_potential}).

Next, we consider the system at finite temperatures and pressures.
\Gls{md} simulations are carried out in the NPT ensemble where the length of the cell vectors are allowed to fluctuate but the angles between them are kept fixed at \qty{90}{\degree}.
The system is cooled at constant pressure from high temperature at a rate of \qty{40}{\kelvin\per\nano\second}, which is sufficiently slow to avoid any noticeable hysteresis.
We also note that it is due to the second-order nature of the phase transition that we can sample and observe it directly in \gls{md} simulations.

To monitor the dynamic evolution of the system we use the temperature dependence of the lattice parameters $a_i$ and the phonon mode coordinates $Q_{\lambda}$.
The latter are obtained by phonon mode projections \cite{SunSheAll2010, FraRosEriRahTadErh23}.
The atomic displacements at each time are scaled back to the original cubic supercell and these scaled displacements $\boldsymbol{u}(t)$ are then projected on a tilt phonon mode $\lambda$ according to
\begin{equation}
    Q_{\lambda}(t) = \boldsymbol{u}(t) \cdot \boldsymbol{e}_{\lambda}\ ,
\end{equation}
where  $\boldsymbol{e}_{\lambda}$ is the supercell eigenvector for mode $\lambda$.
The mode eigenvectors are obtained using \phonopy{} \cite{TogTan15} and symmetrized such that each of the three degenerate modes corresponds to tilting around $x$, $y$, and $z$ directions, respectively.
A gliding time average with width \qty{20}{\ps} is applied along the trajectory of the cooling simulation allowing us to extract the lattice parameters as well as the phonon mode coordinates $Q_{\lambda}$ as a practically continuous function of temperature.

In \autoref{fig:phase_diagram}a the temperature dependence of the lattice parameters and R-mode coordinates are shown at \qty{10}{\GPa} when the system is cooled from \qty{300}{\kelvin}. 
At around \qty{160}{\kelvin} the lattice parameter in the $x$ direction deviates from the other two forming a tetragonal structure at the same time as the R$_x$ mode condensates.
This indicates a phase transition from the cubic $a^0a^0a^0$ ($Pm\bar{3}m$) to the tetragonal $a^0a^0c^-$ ($I4/mcm$) phase.

Cooling runs are repeated for various pressures and the resulting phase diagram is determined and shown in \autoref{fig:phase_diagram}b.
At \qty{300}{\kelvin} we find a phase transition to the tetragonal phase at \qty{16.2}{\GPa}.
We do not see any condensation of the M-tilt modes (in-phase tilting) at any pressure or temperature.
Furthermore, the phase transition only occurs to the tetragonal phase ($I4/mcm$), not to any orthorhombic or rhombohedral phases, except for a small region below \qty{20}{\kelvin} and around 4 to \qty{5}{\GPa}, where the rhombohedral ($R\overline{3}c$) structure becomes stable.
However, for these low temperatures quantum fluctuations become important and we expect these to stabilize the tetragonal structure as discussed below.
The observed lattice parameters as a function of temperature and pressure is shown in \autoref{sfig:lattice_parameters}, and agrees well with experimental work \cite{ZhaoWeidner1991}.

Below about \qty{100}{\kelvin} quantum effects have been shown to be important to correctly model the stability of the cubic phase \cite{RosFraBrauTouBouAndBosMaeWah23}.
Therefore, the phase diagram obtained here using classical \gls{md} simulations becomes less accurate at low temperatures.
This is indicated in \autoref{fig:phase_diagram} by the increased transparency of the color at low temperatures.
We note here that while the classical \gls{md} simulations predict that the system becomes tetragonal at zero temperature and pressure, it is likely not the case if quantum fluctuations are included (see \autoref{sfig:Rmode_frequency} and Ref.~\citenum{RosFraBrauTouBouAndBosMaeWah23}).

The phase transition to the tetragonal phase as function of pressure has been investigated experimentally by Raman spectroscopy \cite{Chemarin2000, Toulouse2022,Gim2022} and \gls{xrd} measurements \cite{Yang2014,Toulouse2022}.
The experimental results are rather scattered.
In Raman studies the phase transition to the tetragonal structure was observed at \qty{11}{\GPa} \cite{Chemarin2000}, \qty{10}{\GPa} \cite{Toulouse2022}, and \qty{19.2}{\GPa} \cite{Gim2022} at room temperature, and in \gls{xrd} measurements at \qty{17.2}{\GPa} \cite{Yang2014} and \qty{12}{\GPa} \cite{Toulouse2022}.
Our observed phase transition, from the cubic to the tetragonal phase, at \qty{16.2}{\GPa} falls approximately in the middle of experimentally observed range. 
In Ref.~\citenum{Gim2022} a transition to a rhombohedral ($R\overline{3}c$) structure was also obtained at \qty{8.4}{\GPa}.
This type of transition was, however, neither observed in the other experimental studies nor does it appear in the present simulations.

\subsection{The structure factor: Temperature dependence}

\begin{figure}[ht]
\centering
\includegraphics{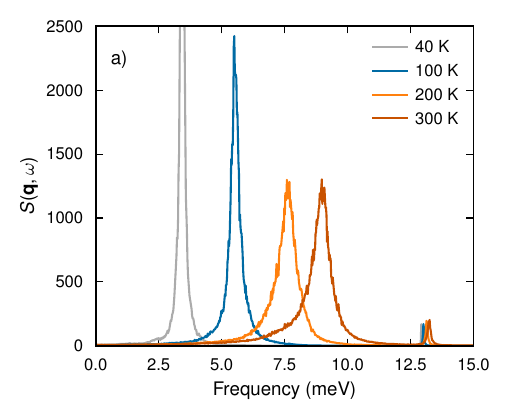}
\caption{
    Dynamical structure factor, $S(\boldsymbol{q}, \omega)$, for various temperatures at \qty{0}{\GPa}.
    Here, $\boldsymbol{q} = \frac{2\pi}{a}(3/2, 1/2, 1/2)$, corresponding to the R-point in the second Brillouin zone.
}
\label{fig:Sqw_temperature}
\end{figure}

Next, we consider the temperature dependence of the structure factor at ambient (zero) pressure.
The intermediate scattering function is defined as 
\begin{align}
    F(\boldsymbol{q},t) = \frac{1}{N} \left < \sum_i^N \sum_j^N \mathrm{exp} \left [ i \boldsymbol{q} \cdot ( \boldsymbol{r}_i(t)-\boldsymbol{r}_j(0)) \right ] \right > \ ,
\end{align}
where $\boldsymbol{r}_i(t)$ denotes the position of atom $i$ at time $t$, $N$ is the number of atoms, and $\left < \ldots \right >$ indicates a time average.
The dynamic structure factor $S(\boldsymbol{q},\omega)$ is obtained by a temporal Fourier transform of $F(\boldsymbol{q},t)$.

We calculate the intermediate scattering function $F(\boldsymbol{q},t)$ from \gls{md} simulations in the NVE ensemble.
The total simulation time for a run is \qty{1}{\ns} and $F(\boldsymbol{q},t)$ is averaged over \num{100} independent such simulations.
The corresponding dynamic structure factor $S(\boldsymbol{q},\omega)$ is shown in \autoref{fig:Sqw_temperature} at the R-point $\boldsymbol{q} = \frac{2\pi}{a}(3/2, 1/2, 1/2)$.

The lower peaks, below \qty{10}{\meV}, correspond to the R tilt-mode and the peaks around \qty{13}{\meV} correspond to the acoustic mode.
The R tilt-mode shows a strong temperature dependence, softening with decreasing temperature.
This is in good agreement with previous experimental and theoretical modeling \cite{RosFraBrauTouBouAndBosMaeWah23, Zheng2022}.

Next, we consider the static structure factor $S(\boldsymbol{q})$, which is related to the intermediate scattering function and the dynamic structure factor via
\begin{align}
    S(\boldsymbol{q}) = F(\boldsymbol{q}, t=0) = \frac{1}{2\pi} \int_{-\infty}^\infty S(\boldsymbol{q}, \omega) \mathrm{d}\omega.
    \label{eq:Sq_Sqw_Fqt}
\end{align}
The partial static structure factors $S_{\alpha\beta}(\boldsymbol{q})$ are then evaluated as
\begin{equation}
    S_{\alpha \beta}(\boldsymbol{q}) = \frac{1}{N} \left < \sum_{i \in \alpha} ^{N_\alpha} \sum_{j \in \beta} ^{N_\beta} \exp{\left [ i\boldsymbol{q}\cdot (\boldsymbol{r}_i(t) - \boldsymbol{r}_j(t)) \right ]} \right >,
\end{equation} 
where $\alpha$ and $\beta$ denote the atom types ($\alpha$~=~Ba, Zr or O), the summation runs over all atoms of the given type and $N_\alpha$ is the number of atoms of type $\alpha$.
The static structure factor is calculated from \gls{md} simulations in the NVT ensemble at \qty{0}{\GPa} and different temperatures.
For each temperature, we average over 40 independent simulations that are each \qty{100}{\ps} long.

\begin{figure}[ht]
\centering
\includegraphics{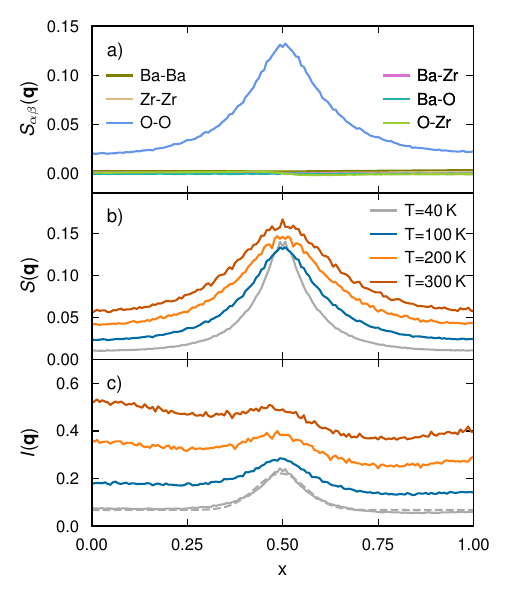}
\caption{
    Structure factors calculated from \gls{md} simulations at \qty{0}{\GPa}.
    Here, $\boldsymbol{q} = \frac{2\pi}{a}(3/2, 1/2, x)$, which corresponds to a horizontal 1D slice of the data given in \autoref{fig:structure_factor_heatmaps}, i.e., starting from an M-point ($x=0$), passing an R-point ($x=1/2$), and ending at another M-point ($x=1$).
    a) Partial structure factors $S_{\alpha\beta}(\boldsymbol{q})$ at \qty{100}{\kelvin}.
    b) Total structure factor $S(\boldsymbol{q})$ at 40, 100, 200 and \qty{300}{\kelvin}.
    c) Scattering intensity $I(\boldsymbol{q})$ at 40, 100, 200 and \qty{300}{\kelvin}. The dashed line for \qty{40}{\kelvin} shows a Gaussian fit.
    }
\label{fig:structure_factor_slice}
\end{figure}

We consider first the partial static structure factors $S_{\alpha\beta}$ at \qty{100}{\kelvin} calculated along the Brillouin zone path $\boldsymbol{q} = \frac{2\pi}{a}(3/2, 1/2, x)$ with $x: 0 \to 1$, shown in \autoref{fig:structure_factor_slice}a, corresponding to a path $M \to R \to M$ also used by Levin \textit{et al.} \cite{Levin2021} (see their Fig.~5).
The oxygen-oxygen part gives rise to large intensity at the R-point ($x=1/2$), in agreement with the soft oxygen tilt mode at R, as well as a background intensity.
The other partial static structure factors only give rise to a very weak background intensity.

The temperature dependence of the static structure factor $S(\boldsymbol{q})$ is shown in \autoref{fig:structure_factor_slice}b at 40, 100, 200, and \qty{300}{\kelvin}.
For all temperatures there is a peak at the R-point ($x$=1/2).
To further understand this, consider the static structure factor for a harmonic system \cite{AshMer76} in the classical limit
\begin{equation}
    S(\boldsymbol{q}) \propto \sum_\lambda^{N_\text{modes}} |F^\text{ph}_\lambda(\boldsymbol{q})|^2 \frac{2 k_\text{B}T}{(\hbar \omega_\lambda)^2}\ ,
    \label{eq:harmonic_Sq}
\end{equation}
where the sum runs over all phonon modes for the given $\boldsymbol{q}$-point and $F^\text{ph}_\lambda(\boldsymbol{q})$ is the phonon structure factor containing the Debye-Waller factor and mode selection rules \cite{Chou2000}.
Therefore, we roughly expect the intensity to increase linearly with temperature $T$ and to scale with frequency as $1/\omega^2$.
The peak height of $S(\boldsymbol{q})$ is almost constant with temperature whereas the background increases linearly with temperature in accordance with a harmonic system. 
The constant peak height is due to that the tilt-frequency of the R-mode softens from \qty{9}{\meV} to about \qty{3}{\meV} between \qty{300}{\kelvin} and \qty{40}{\kelvin}, since $300/9^2 \approx 40/3^2$. Thus, the structure factor at the R-point remains more or less constant with temperature.
The clear peak at \qty{40}{\kelvin} is therefore a result of the tilt-frequency of the R-mode softening with temperature.

The present \gls{md} simulations are based on classical mechanics.
Quantum fluctuations of the atomic motion start to become important for the R-mode frequency below \qty{100}{\kelvin} \cite{RosFraBrauTouBouAndBosMaeWah23}.
We have tested the effect quantum fluctuations on the above peak height using a self-consistent phonon approach (see \autoref{sfig:Sq_harmonic}).
The peak height at the R-point is slightly reduced by including the quantum effects, but a clear peak at \qty{40}{\kelvin} is still present.

Lastly, to get a one-to-one comparison with the electron beam diffraction experiments carried out by Levin \textit{et al.} \cite{Levin2021}, we determine the intensity, $I(\boldsymbol{q})$, by weighting the partial structure factors with their corresponding electron atomic scattering factors according to
\begin{equation}
    I(\boldsymbol{q}) = \sum_\alpha \sum_\beta f_\alpha(q) f_\beta(q) S_{\alpha \beta}(\boldsymbol{q}) \ .
\end{equation}
Here, $f_\alpha(q)$ are the $q$-dependent electronic scattering factors for the ions \ce{Ba^{2+}}, \ce{Zr^{4+}}, and \ce{O^{2-}}, with numerical data taken from Ref.~\citenum{Pen98} (see \autoref{sfig:scattering_factors}).
The scattering factors are roughly proportional to the atomic number, reducing the oxygen contribution significantly.
The peak at the R-point for the intensity $I(\boldsymbol{q})$ is reduced in height (relative to the background) (\autoref{fig:structure_factor_slice}c) and there is barely any visible peak above \qty{100}{\kelvin}. 
This is in good agreement with the observation by Levin \textit{et al.} \cite{Levin2021} that a weak and diffuse, yet discrete spot appears at the R-point below about \qty{80}{\kelvin}.
For \qty{40}{\kelvin} we also carry out a Gaussian fit of $I(\boldsymbol{q})$ with a constant background to extract the \gls{fwhm} of \qty{0.23}{\per\angstrom}, which is in very good agreement with the value of \qty{0.22}{\per\angstrom}, reported by Levin \textit{et al.} \cite{Levin2021}.

\begin{figure}[ht]
\centering
\includegraphics{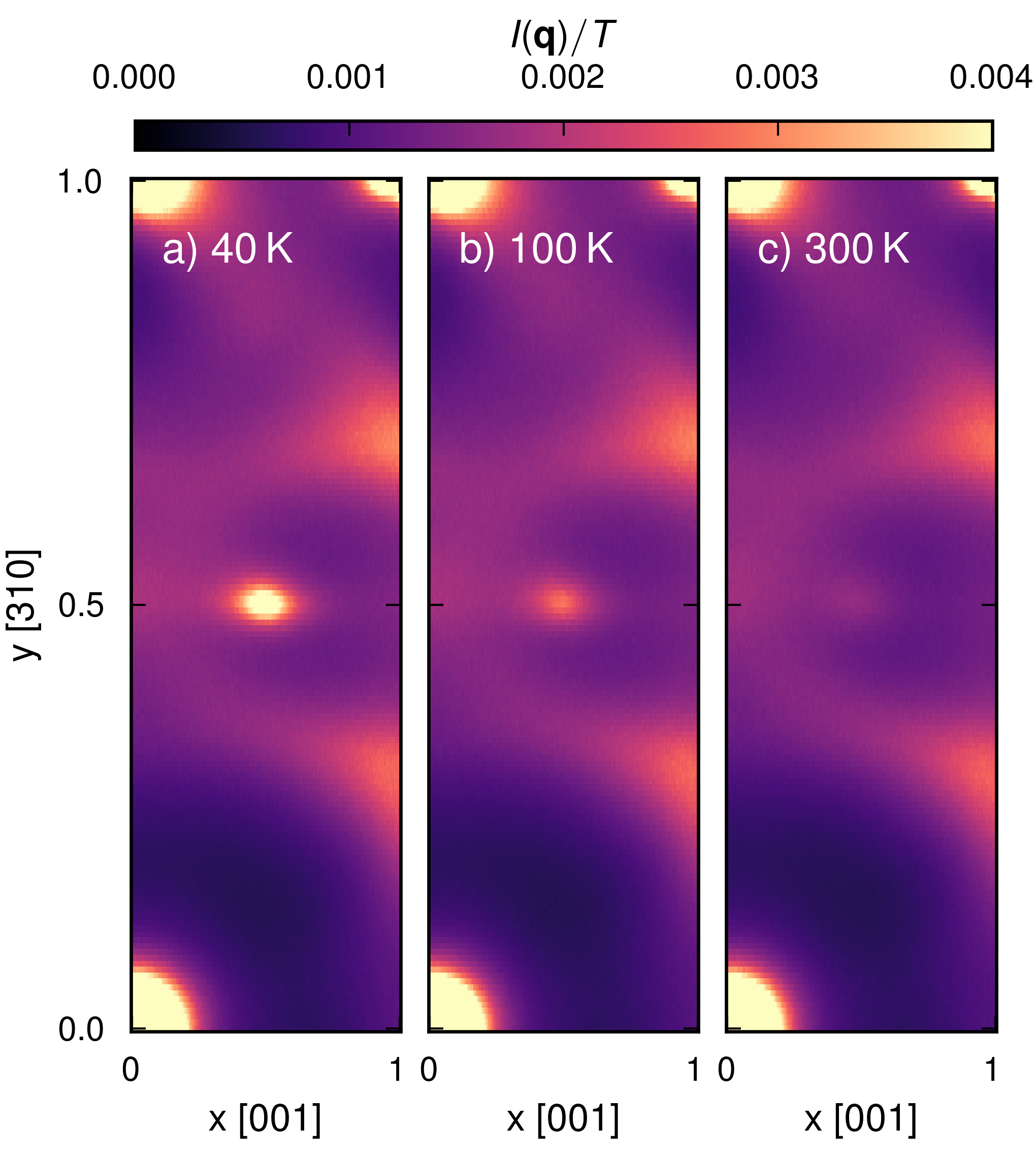}
\caption{
    Intensity normalized by temperature, $I(\boldsymbol{q})/T$, calculated from \gls{md} simulations at a) \qty{40}{\kelvin}, b) \qty{100}{\kelvin}, and c) \qty{300}{\kelvin}.
    Here, $\boldsymbol{q} = \frac{2\pi}{a}(3y, y, x)$.
    The lower left corner thus corresponds to the $\vec{q}$-point $\boldsymbol{q}=(0,0,0)$, the upper left corner to $\boldsymbol{q}=\frac{2\pi}{a} (3,1,0)$, the lower right corner to $\boldsymbol{q}=\frac{2\pi}{a} (0,0,1)$, and the upper right corner to $\boldsymbol{q}=\frac{2\pi}{a} (3,1,1)$.
    The center of the heatmaps corresponds to the R-point $\boldsymbol{q}=\frac{2\pi}{a}(3/2,1/2,1/2)$.
    Note also that the color scale is set such that diffuse scattering is visible, but in practice the intensity at the $\Gamma$-points are orders of magnitude larger.
}
\label{fig:structure_factor_heatmaps}
\end{figure}

Next, we extend the calculation of the intensity $I(\boldsymbol{q})$ to the same 2D space of $\vec{q}$-points as highlighted by Levin \textit{et al.} \cite{Levin2021} in their Figure~4.
Because the intensity increases almost linearly with temperature (Eq.~\eqref{eq:harmonic_Sq}), we plot $I(\boldsymbol{q})/T$ to enable easier comparison between temperatures.
These normalized intensities are shown as heatmaps in \autoref{fig:structure_factor_heatmaps}.
Heatmaps for the partial intensities $I_{\alpha\beta}(\boldsymbol{q})$ at \qty{100}{\kelvin} can be found in \autoref{sfig:partial_heatmaps}.
Most of the intensity heatmaps in \autoref{fig:structure_factor_heatmaps} look very similar for all three temperatures.
The larger intensities in the corners ($\Gamma$ points) corresponds to the Bragg peaks.
The intensity between Bragg peaks, the diffuse scattering, arises due to thermal motion.
The only real notable difference between the temperatures is the increased intensity in the middle of the heatmap (at the R-point $\boldsymbol{q}=\frac{2\pi}{a}(3/2,1/2,1/2)$) for lower temperatures.
At \qty{300}{\kelvin} there is almost no peak visible at the R-point compared to the intensity level for the surrounding $\boldsymbol{q}$-points, whereas for \qty{40}{\kelvin} there is a very clear peak (as also seen in \autoref{fig:structure_factor_slice}c).
The notable intensities at $x$=1, $y$=1/3 and $x$=1, $y$=2/3 arise from the low frequency Ba and Zr modes between $\Gamma$ and M (and close to X) in the phonon dispersion ($\it cf.$ \autoref{sfig:dispersion_along_311} and \autoref{sfig:partial_heatmaps}).

\subsection{Tilt angle correlations: Temperature dependence}

\begin{figure*}[ht]
\centering
\includegraphics{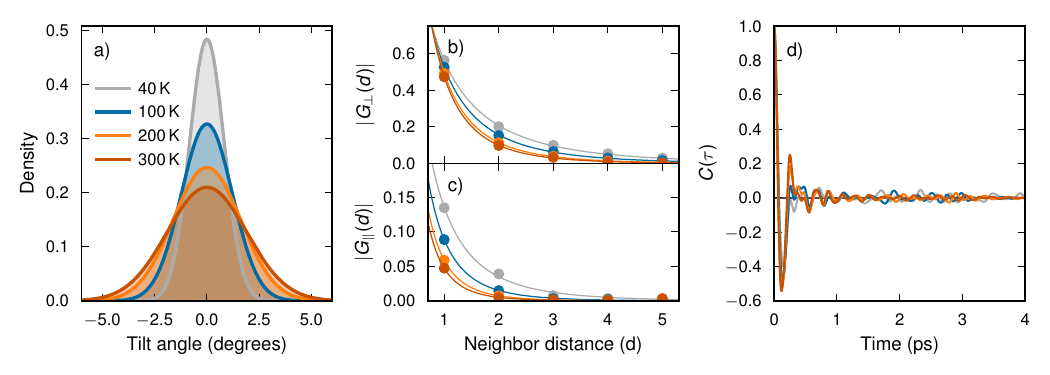}
\caption{
    Distributions of the tilt angle $\theta^{\alpha}$ and its correlations from \gls{md} simulations at \qty{0}{\GPa} and different temperatures.
    a) Tilt angle distribution, averaged over $\alpha=x,y,z$.
    Solid lines represent Gaussian fits with zero mean.
    b,c) Static tilt-angle correlation functions $G_\parallel(d)$ and  $G_\perp(d)$ as a function of neighbor distance $d$.
    Solid lines are guides to the eye.
    d) Dynamic tilt-angle correlation function $C(\tau)$ as defined in Eq.~\eqref{eq:acf} and averaged over $\alpha=x,y,z$.
}
\label{fig:local_tilts_temperature}
\end{figure*}

To obtain a more local picture we now consider the tilt angle of each individual \ce{ZrO6} octahedron, and its static and dynamic correlations.
We first extract the Euler angles for each octahedron from \gls{md} simulations.
We employ the polyhedral template matching using \ovito{} \cite{Stukowski2010} as done in Ref.~\citenum{WikFraErh2023}.
This allows us to extract tilt angles around the $\alpha$ axis ($\alpha = x, y, z$) for an octahedron located at $(n_x, n_y, n_z)$ at time $t$, $\theta^\alpha(n_x, n_y, n_z, t)$.
Here, we follow a similar notation as in Refs.~\citenum{Baldwin2023, Liang2023}.

The distribution $P(\theta)$ over $\theta^\alpha(n_x, n_y, n_z, t)$ averaged over $\alpha$, $(n_x,n_y,n_z)$, and $t$ can now be determined (\autoref{fig:local_tilts_temperature}a).
We notice that for all temperatures the distribution exhibits a Gaussian profile with zero mean and with a standard deviation that increases with temperature.
The standard deviations are $\sigma=\qty{0.82}{\degree}$, $\sigma=\qty{1.21}{\degree}$, $\sigma=\qty{1.62}{\degree}$, and $\sigma=\qty{1.90}{\degree}$ for \qty{40}{\kelvin}, \qty{100}{\kelvin}, \qty{200}{\kelvin}, and \qty{300}{\kelvin}, respectively.
In a classical harmonic system we expect the variance $\sigma^2$ to increase linearly with temperature but here, due to the softening of the R-tilt mode, the distribution over tilt angles shows a weaker temperature dependence.

Next, we consider the static tilt-angle correlation function between $\theta^\alpha(n_x, n_y, n_z, t)$ and its neighboring octahedra.
Here, we only consider neighbors along the [100], [010] and [001] directions.
The static correlation function in the [100] direction is calculated as
\begin{align*}
G_x^\alpha(d) = \frac{\left <  \theta^\alpha(n_x+d, n_y, n_z, t) \theta^\alpha(n_x, n_y, n_z, t)\right >}{\left < \theta^\alpha(n_x, n_y, n_z, t) \theta^\alpha(n_x, n_y, n_z, t) 
\right >}\ ,
\end{align*}
where $d$ corresponds to the number of neighbor distances between two octahedra in the $x$-direction and $\left < \ldots \right >$ denotes an average carried out over $(n_x,n_y,n_z)$ and $t$.
Similarly, one can define the static correlation function along the [010] direction, $G_y^\alpha(d)$, and along the [001] direction, $G_z^\alpha(d)$.
In the cubic phase we only obtain two symmetrically distinct  static correlation functions, $G_\perp(d)$ and  $G_\parallel(d)$, corresponding to if the rotation axis (superscript $\alpha$) is perpendicular or parallel to the neighbor direction, respectively.

The result for the static tilt-angle correlation functions are shown in \autoref{fig:local_tilts_temperature}b.
For both $G_\perp(d)$ and $G_\parallel(d)$ the correlation alternates between positive and negative values when increasing the neighbor distance, since the R-tilt mode dominates that motion \cite{Levin2021, Baldwin2023, WikFraErh2023}.
We thus only show $|G_\perp(d)|$ and $|G_\parallel(d)|$ in \autoref{fig:local_tilts_temperature}b.
The alternation of the correlation is demonstrated in \autoref{sfig:tilt_angle_2D_distribution}, where the joint probability distribution over two angles is shown.
For the correlation perpendicular to the rotational axis, $|G_\perp(d)|$, we find a strong correlation between nearest neighbor octahedra which decays towards zero after about 4 to 5 neighbor distances, corresponding to about \qty{2}{nm} (\autoref{fig:local_tilts_temperature}b).
In the direction parallel to the rotation axis, the correlation is weaker and decays faster.
This is related to the soft phonon mode at the M-point corresponding to in-phase tilting of the octahedra that thus to some extent counteracts the out-of-phase tilting by the R-mode.
For both $G_\perp(d)$ and $G_\parallel(d)$ we see that the correlation increases with decreasing temperature.
This is connected to the softening of the R-mode frequency which causes the correlation length to increase.

Lastly, we consider the dynamic autocorrelation function for the tilt angles $\theta^\alpha(n_x, n_y, n_z, t)$ defined as
\begin{align}
    C^\alpha(\tau) = \frac{\left <  \theta^\alpha(n_x, n_y, n_z, t+\tau) \theta^\alpha(n_x, n_y, n_z, t)\right >}{\left < \theta^\alpha(n_x, n_y, n_z, t) \theta^\alpha(n_x, n_y, n_z, t) \right >}\ ,
    \label{eq:acf}
\end{align}
where $\left < \ldots \right >$ corresponds to an averaged carried out over $(n_x,n_y,n_z)$ and $t$.
The result for the correlation function, averaged over $\alpha$, is shown in \autoref{fig:local_tilts_temperature}c.
For all four temperatures the correlation function oscillates at short-time scales and then goes to zero after a few picoseconds.
This is clear indication that there are no static or "frozen in" tilts for the temperatures considered, but rather the tilts are dynamically changing on a picosecond time-scale.

\subsection{The structure factor - pressure dependency}

Let us now consider the pressure dependence of the dynamic and static structure factors at \qty{300}{\kelvin}.
The system is studied from \qty{0}{\GPa} to \qty{18}{\GPa} and at \qty{16.2}{\GPa} it transforms from the cubic to the tetragonal phase.

\begin{figure}[ht]
\centering
\includegraphics{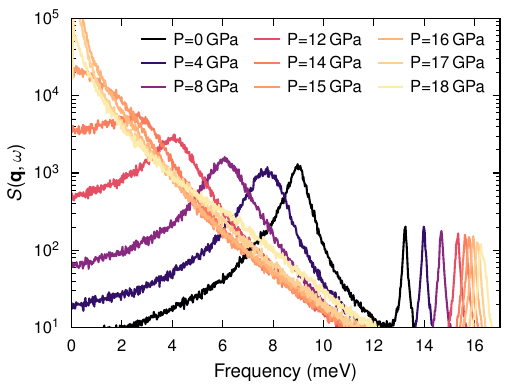}
\caption{
    The dynamical structure factor, $S(\boldsymbol{q},\omega)$, calculated from \gls{md} for various pressures at \qty{300}{\kelvin}.
    Here, $\boldsymbol{q} = \frac{2\pi}{a}(3/2, 1/2, 1/2)$.
}
\label{fig:Sqw_pressure}
\end{figure}

The pressure dependence for the dynamic structure factor is shown in \autoref{fig:Sqw_pressure}.
For \qty{0}{\GPa} we see a clear peak at around \qty{9}{\meV} corresponding to the R-tilt mode.
The peaks above \qty{12}{\meV} correspond to the acoustic mode.
The frequency of the R-tilt mode decreases with increasing pressure and the magnitude of the dynamic structure factor increases substantially (Notice the logarithmic scale on the y-axis.)
At the same time the damping of the mode increases and at around \qty{15}{\GPa} it becomes overdamped.

\begin{figure}[ht]
\centering
\includegraphics{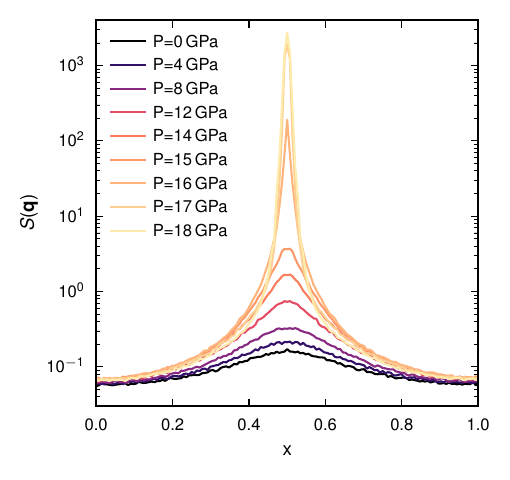}
\caption{
    The structure factors, $S(\boldsymbol{q})$, calculated from \gls{md} for various pressures at \qty{300}{\kelvin}.
    Here, $\boldsymbol{q} = \frac{2\pi}{a}(3/2, 1/2, x)$
}
\label{fig:structure_factor_slice_pressure}
\end{figure}

The corresponding static structure factor is shown in \autoref{fig:structure_factor_slice_pressure}.
The static structure factor has the shape of a Lorentzian peak on a log-scale. 
The width of peak decreases when approaching the phase transition, indicating that the correlation length increases.
Further, the value of static structure factor increases exponentially at the R-point as one approaches the phase transition pressure.
This can be understood from the fact that the frequency of the R-tilt mode approaches zero at the phase transition and thus $S(\boldsymbol{q})$ diverges (cf.~\autoref{eq:harmonic_Sq}).
Furthermore, the large values of $S(\boldsymbol{q}, \omega)$ observed for higher pressures at low frequencies is directly related to the divergence of the static structure factor $S(\boldsymbol{q})$ (\autoref{fig:structure_factor_slice_pressure}), as can be seen from \autoref{eq:Sq_Sqw_Fqt}.

\begin{figure*}[ht]
\centering
\includegraphics{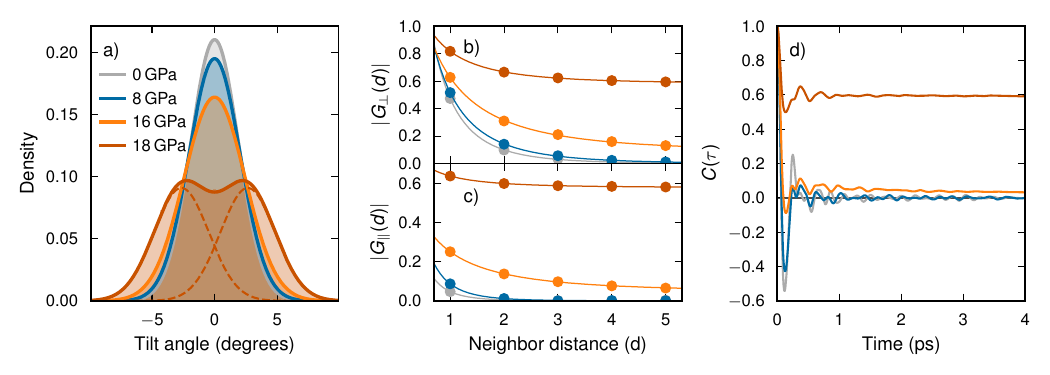}
\caption{
    Properties of the tilt-angle $\theta^{\alpha}$ and its static  and dynamic correlations from \gls{md} simulations at \qty{300}{\kelvin} and for four different pressures.
    The tilt-angle distribution and its correlations are averaged over $\alpha=x,y,z$ for \num{0}, \num{8}, and \qty{16}{\GPa}.
    For \qty{18}{\GPa} they are only calculated over the direction $\alpha$ for which the tetragonal structure is tilted around.
    a) Distribution over observed tilt-angles.
    Solid lines corresponds to Gaussian fits with zero mean, or symmetric Gaussians with nonzero mean for \qty{18}{\GPa}.
    The dashed lines correspond to the two symmetric Gaussians for \qty{18}{\GPa}.
    b) and c) The static tilt-angle correlation functions, $G_\parallel(d)$ and  $G_\perp(d)$, as a function of neighbor distance $d$.
    Solid lines are guides to the eye.
    c) The dynamic tilt-angle correlation function $C(\tau)$ as defined in Eq.~\eqref{eq:acf}.
}
\label{fig:local_tilts_pressure}
\end{figure*}

\subsection{Tilt angle correlations - pressure dependency}

Finally, we consider the tilt angle and its static and dynamic correlations as function of pressure, shown in \autoref{fig:local_tilts_pressure}.
The highest pressure, \qty{18}{\GPa}, is located above \qty{16.2}{\GPa}, the pressure where the system transform from the cubic to the tetragonal phase. 
The data for \qty{18}{\GPa} is therefore calculated using only the direction $\alpha$ for which the tetragonal structure is tilted around. 
For the three lower pressures the data are obtained by making an average of the three different $\alpha$ directions.

The distribution for the tilt angle $P(\theta)$ as function of pressure is shown in \autoref{fig:local_tilts_pressure}a.
The distribution widens with increasing pressure.
For \qty{18}{\GPa}, where the system has undergone the phase transition to the tetragonal phase, the distribution develops a symmetric double peak distribution.
This can be fitted well with two Gaussians with mean values $\mu_t=\pm$\qty{2.65}{\degree} and standard deviation $\sigma_t=\ $\qty{2.19}{\degree}.

The static tilt-angle correlation function as a function of neighbor distance is shown in \autoref{fig:local_tilts_pressure}b.
The static correlations increase as function of pressure and the decay distance increases.
Above the phase transition the correlations do not decay to zero and the correlation function approaches the constant value
\begin{align*}
 |G(d\rightarrow\infty)| = 
 \frac{\left< \theta_t \right> \left< \theta_t \right>}
 {\left< \theta_t^2 \right>} =
 \frac{\mu_t^2}{\mu_t^2 + \sigma_t^2} =
 0.59
\end{align*}
for \qty{18}{\GPa}, reflecting the (global) long-ranged tilting in the tetragonal phase.

Similar behavior is also seen in the dynamic tilt-angle autocorrelation function $C^{\alpha}(\tau)$ in \autoref{fig:local_tilts_pressure}c.
For pressures below the phase transitions $C^{\alpha}(\tau)$ decays to zero in the long-time limit, whereas for \qty{18}{\GPa} $C^{\alpha}(\tau)$ approaches the same constant value as the static correlation function, i.e., $C^{\alpha}(\tau\rightarrow\infty) = 0.59$.
It is interesting to note that just below the phase transition the decay time increases substantially.
This has also been seen in similar simulation studies of halide perovskites \cite{FraRosEriRahTadErh23, Baldwin2023}.


\section{Discussion}


Structural instabilities and phase transitions in perovskite oxides are important and have therefore been investigated extensively.
Strontium titanate \ce{SrTiO3} (STO) is generally considered to be a model perovskite for the study of soft mode-driven phase transitions \cite{Fleury1968, Cowley1996} and it may be instructive to compare the behavior of STO with BZO.

At ambient conditions STO is cubic and its antiferrodistortive transition to the tetragonal ($I4/mcm$) phase can be induced by either decreasing the temperature or increasing the pressure \cite{Weng2014}. 
The pressure induced transition at room temperature occurs at \qty{9.6}{\GPa} for STO \cite{Guennou2010}.
The same type of transition also occurs in BZO but at a somewhat higher pressure \cite{Yang2014, Toulouse2022}.
On the other hand, the temperature induced transition at ambient pressure only occurs in STO, not in BZO.
In STO the R-tilt mode softens and at about \qty{105}{\kelvin} \cite{Fleury1968, Holt2007} it approaches zero and the material undergoes a phase transition to the tetragonal structure.
When approaching this phase transition from above the scattering intensity near the R-point increases dramatically and the scattering peak narrows substantially in $\vec{q}$-space \cite{Holt2007}.

Our results demonstrate that a similar mechanism is also at play in BZO and detected in the experiments by Levin \textit{et al.} \cite{Levin2021}.
Yet in contrast to STO, one only reaches the initial narrowing of the peak as the phase transition never occurs at ambient pressure.
The R-tilt mode softens but its frequency remains finite when the temperature goes to zero \cite{Zheng2022, RosFraBrauTouBouAndBosMaeWah23}.
Levin \textit{et al.} \cite{Levin2021} find a diffuse peak at the R-point with a width of \qty{0.22}{\per\angstrom}.
This magnitude corresponds roughly to the corresponding peak for STO at about \qty{160}{\kelvin}, that is \qty{50}{\kelvin} above the transition to the tetragonal phase \cite{Holt2007}.
The ``nanodomains'' observed by Levin \textit{et al.} \cite{Levin2021} are thus dynamic correlations at the onset of a phase transition that never occurs in BZO at ambient pressures.

\section{Conclusions}

We have performed large scale \gls{md} simulations of barium zirconate, an oxide perovskite, using machine-learned potentials based on \gls{dft} calculations.
Both the temperature and pressure dependence of the local and global structure and the dynamics were investigated, and compared with available electron diffraction results.

At ambient pressure it is now well established that BZO remains cubic down to zero Kelvin, although the R-tilt mode softens substantially \cite{RosFraBrauTouBouAndBosMaeWah23}.
Our \gls{md} simulations predict a softening from \qty{9}{\meV} at \qty{300}{\kelvin} to \qty{3}{\meV} at \qty{40}{\kelvin}.
We find that this mode softening gives rise to a clear oxygen related peak in the static structure factor at the R-point, which explains the superlattice reflection observed by Levin \textit{et al.} \cite{Levin2021} using electron diffraction.

Levin \textit{et al.} \cite{Levin2021} state that the peak is only visible below about \qty{80}{\kelvin}. 
However, we show that it does exist also at higher temperatures, albeit with weaker intensity.
The present study strongly suggests that the disappearance of the peak in the electron diffraction study at higher temperatures is due to the large background intensity from scattering of Ba and Zr at those temperatures.
The oxygen related peak is the result of strongly correlated and dynamic tilting between neighboring \ce{ZrO6} octahedra.
By investigating the tilt angle correlations we find that the spatial extent of the correlated motion at \qty{40}{\kelvin} is about 2 to \qty{3}{\nm} and with a short relaxation time of about \qty{1}{\ps}. 
We therefore conclude that the oxygen peak observed at the R-point is purely of dynamic origin.

The pressure dependence at room temperature was also investigated.
It is known that BZO undergoes a phase transition from the cubic to the tetragonal phase. 
Here, we obtain this transition at about \qty{16}{\GPa} in the middle of the experimentally observed range.
When approaching the phase transition from lower pressures, the frequency of the R-tilt mode approaches zero and close to the phase transition the motion becomes overdamped. 
At the same time the static structure factor at the R-point increases dramatically by several orders of magnitude. 
The dynamic tilt-angle autocorrelation function shows a rapid decay on the order of \qty{1}{\ps}, but close to the transition, the correlation function also develops a component with a considerably slower decay.
At the phase transition this decay goes over to a constant finite value. 
The static tilt-angle correlation function shows a similar behavior: The decay rate becomes longer and longer and the correlation function approaches a constant value at the phase transition.

The present study shows that large scale \gls{md} simulations based on a machine-learned potentials with near \gls{dft} accuracy can provide immensely detailed and accurate atomic scale information on the local structure and complex dynamics close to phase transitions.


\section{Methods}

\subsection{Reference calculations}

The energy, forces, and virials are obtained for the training structures via \gls{dft} calculations as implemented in the Vienna ab-initio simulation package \cite{KreHaf93, KreFur1996-1, KreFur1996-2} using the projector-augmented wave \cite{Blo94,KreJou99} setups in version 5.4.4 with a plane wave energy cutoff of \qty{510}{\electronvolt}.
The considered valence configurations for Ba, Zr and O are $5s^25p^66s^2$, $4s^24p^64d^25s^2$ and $2s^22p^2$, respectively.
The Brillouin zone is sampled with a Monkhorst-Pack grid, with the maximum distance between two points being \qty{0.19}{\per\angstrom} along the reciprocal lattice vectors.
This leads to a \numproduct{8x8x8} $\vec{k}$-point grid for the primitive cell with a lattice parameter of \qty{4.20}{\angstrom}.

For the exchange-correlation functional we employ the van-der-Waals-density functional with consistent exchange (vdW-DF-cx) \cite{DioRydSch04, BerHyl2014}, here abbreviated CX.
This functional is a version of the vdW-DF method \cite{Berland2015},
with the aim of accurately capturing competing interactions in soft and hard materials \cite{BerArtCooLeeLunSchThoHyl14, Frostenson2022}.
It has been applied to \ce{BaZrO3} before \cite{PerJedRomPioLinHylWah20, JedWahHyl20} and been found to give a very good account of its structural and vibrational properties.
In particular, the anharmonicity of the R-tilt mode at ambient pressure is well described compared to recent experiments on \ce{BaZrO3} \cite{RosFraBrauTouBouAndBosMaeWah23}, and so is the thermal expansion \cite{JedWahHyl20} (for more details see \autoref{sfig:lattice_parameters} and \autoref{sfig:Rmode_frequency_XC}).

\subsection{Neuroevolution potential}

We construct an \gls{nep} model for the potential energy surface using the iterative strategy outlined in Ref.~\citenum{FraWikErh2023}.
Training structures include cubic, tetragonal and rhombohedral primitive cells at different volumes and cell-shapes, \gls{md} structures in a \numproduct{4x4x4} (320 atoms) supercell at temperatures up to \qty{500}{\kelvin} and pressures up to \qty{40}{\giga\pascal}, structures with various tilt-modes imposed, cubic-tetragonal and tetragonal-tetragonal interface structures, and structures found by simulated annealing at different pressures.
The \gls{md} structures are generated using an initial \gls{nep} model and are selected based on their uncertainty, which is estimated from the predictions of an ensemble of models \cite{FraWikErh2023}.
The final \gls{nep} model used in the production runs is trained on all the available training data (see \autoref{sfig:active_learning}).
The \gls{nep} model accurately reproduces the energy-volume curves for the different phases, the phonon dispersions for the cubic phase as well as the static energy landscape of the tilt modes (R and M).
More details pertaining to the validation of the \gls{nep} model including parity plots are provided in the Supporting Information.

\subsection{Molecular dynamics}

All \gls{md} simulations are run with \gpumd{} \cite{Fan17}.
In all simulations we employ a timestep of \qty{1}{\femto\second} and equilibration time of \qty{50}{\pico\second}.
For most simulations we use supercells comprising \numproduct{24x24x24} cubic primitive cells ($\sim$ 70 thousand atoms).
However, the static structure factor $S(\boldsymbol{q})$ is calculated from \gls{md} simulations with \numproduct{144x144x144} cubic primitive cells ($\sim$ 15 million atoms)
in order to achieve an adequate $\vec{q}$-point resolution.
The static and dynamic structure factors are calculated from \gls{md} trajectories using the \dynasor{} package \cite{FraSlaErhWah2021}.

The phase diagram is obtained from simulations in the NPT ensemble, static properties from simulations in the NVT ensemble, and dynamic properties from simulations in the NVE ensemble.
NVT and NVE simulations are carried out with lattice parameters obtained from NPT simulations (see \autoref{sfig:lattice_parameters}).

\section*{Acknowledgments}
Funding from the Swedish Energy Agency (grant No. 45410-1), the Swedish Research Council (2018-06482, 2020-04935, and 2021-05072), the Area of Advance Nano at Chalmers, and the Chalmers Initiative for Advancement of Neutron and Synchrotron Techniques is gratefully acknowledged. 
The computations were enabled by resources provided by the National Academic Infrastructure for Supercomputing in Sweden (NAISS) at PDC, C3SE, and NSC, partially funded by the Swedish Research Council through grant agreement no. 2022-06725.
Computational resources provided by Chalmers e-commons are also acknowledged.

\section*{Data Availability}
The \gls{nep} model for \ce{BaZrO3} constructed in this study as well as a database with the underlying \gls{dft} calculations is openly available via Zenodo at \url{https://doi.org/10.5281/zenodo.8337182}.

\section*{Supporting information}
The supporting information provides details pertaining to the \gls{dft} calculations, the \gls{nep} construction and the validation of the \gls{nep}.
Furthermore, the supporting information contains additional results and figures like thermal expansion, static structure factors, and more details on the local-tilt angles.

\end{document}